\pdfoutput=1
\documentclass{PoS}
\usepackage{amsmath}
\usepackage{amsfonts}
\usepackage{amssymb}
\usepackage{mathbbol}
\usepackage{wrapfig}
\usepackage{subfig}
\usepackage{float}
\newcommand{\Projector}{\ensuremath{\mathcal{P}}}
\newcommand{\B}{B_\Projector^{C\gamma_5}}

\newcommand{\qdoty}{\vec{q}\cdot \vec{y}}
\newcommand{\tsink}{\ensuremath{t_{\mathrm{sink}}}}
\newcommand{\tins}{\ensuremath{t_{\mathrm{ins}}}}
\newcommand{\volfac}{\ensuremath{a^3}}
\renewcommand{\Eins}{\mathbb{1}}

\newcommand{\Psibar}{\overline{\Psi}}

\newcommand{\Tr}{\mathrm{tr}}
\renewcommand{\i}{\mathrm{i}}
\newcommand{\e}{\mathrm{e}}

\def\eqref#1{Eq.~(\ref{#1})}

\def\fig#1{Fig.~\ref{#1}}

\title{Nucleon structure from stochastic estimators}
\ShortTitle{Nucleon structure from stochastic estimators}

\author{Gunnar S. Bali, Sara Collins, Benjamin Gl\"a\ss{}le, Meinulf G\"ockeler, \speaker{Johannes Najjar}, Rudolf R\"odl, Andreas Sch\"afer, Andr\'e Sternbeck, Wolfgang S\"oldner \\
        University of Regensburg, Universit\"atsstra\ss{}e 31, 93053 Regensburg\\
        E-mail: \email{Gunnar.Bali@ur.de}, \email{Sara.Collins@ur.de}, \email{Benjamin.Glaessle@ur.de}, \email{Meinulf.Goeckeler@ur.de}, \email{Johannes.Najjar@ur.de}, \email{Rudolf.Roedl@ur.de}, \email{Andreas.Schaefer@ur.de}, \email{Andre.Sternbeck@ur.de},   \email{Wolfgang.Soeldner@ur.de}}

\abstract{Using stochastic estimators for connected meson and baryon three-point functions has successfully been tried in the past years. Compared to the standard sequential source method we trade the freedom to compute the current-to-sink propagator independently of the hadron sink for additional stochastic noise in our observables. In the case of the nucleon we can use this freedom to compute many different sink-momentum/polarization combinations, which grants access to more virtualities. We will present preliminary results on the scalar, electro-magnetic and axial form factors of the nucleon in $N_f=2+1$ lattice QCD and contrast the performance of the stochastic method to the sequential source method. We find the stochastic method to be competitive in terms of errors at fixed cost.}

\FullConference{31st International Symposium on Lattice Field Theory - LATTICE 2013\\
		July 29 - August 3, 2013\\
		Mainz, Germany}

\begin{document}

\section{Introduction}
The structure of hadrons can be parameterized by generalized parton distribution functions. Their moments are accessible via three-point functions on the lattice.

\begin{figure}[!h]
\begin{center}
\subfloat[Sequential Propagator]{\includegraphics[width=0.22\textwidth]{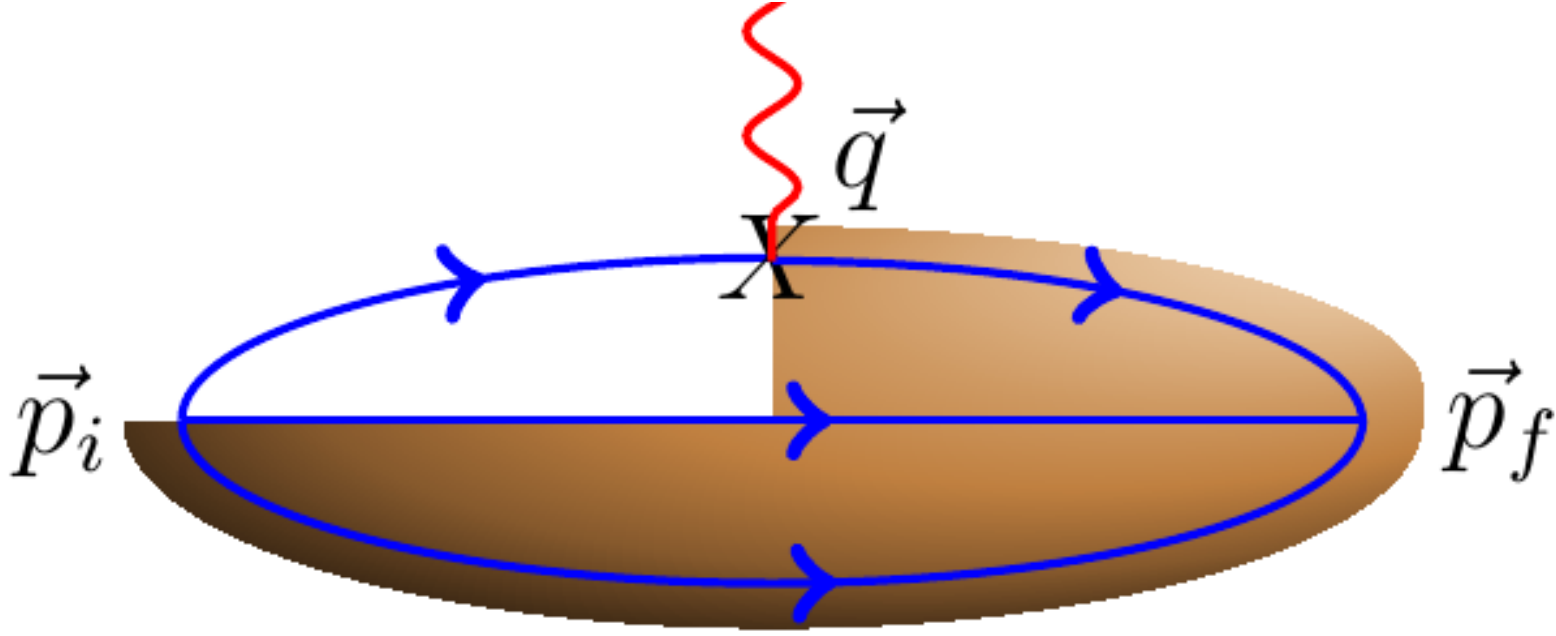}}%
\hspace{0.05\textwidth}\subfloat[Stochastic estimation: The insertion and sink dependence factorize\label{pic:Comicb}]{\includegraphics[width=0.7\textwidth]{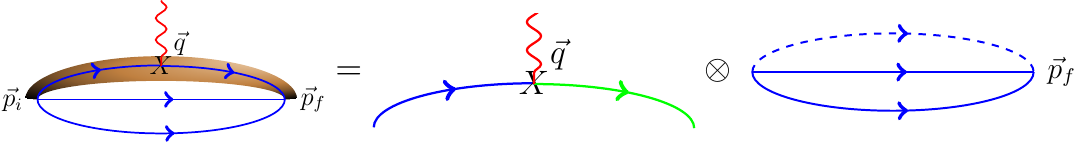}}
\caption{Pictorial representation of connected baryon three-point functions.\label{pic:Comic}}
\end{center}
\end{figure}
Three-point functions are traditionally computed with the sequential source method \cite{Martinelli:1988rr}  and in this work we present an alternative approach using stochastic estimation techniques.

The advantage of this stochastic method \cite{Evans2010} for baryon three-point functions \cite{Alexandrou2013} is that it allows for all polarizations and many sink momenta to be computed simultaneously, as both the current and the sink dependence factorize, see \fig{pic:Comicb}. 
The latter grants access to high momentum transfers with small momenta at the source and sink, as the transfered momentum can be split between the baryon source and sink. 
Thus, two-point functions at only small momenta are needed. 
As the signal for the two-point functions deteriorates faster with increasing momentum, this could potentially outweigh the additional noise from the stochastic estimation. 
To determine this we performed a benchmark analysis of form factors with the stochastic method. 

\section{Computation of three-point functions with stochastic estimates}
The two-point function for the nucleon is given by
\begin{eqnarray}
  C^{\overline{N}N}_{\Projector}( \vec{p}, t) & =& V_3   \volfac\sum_{\vec{x} \in V_3} \e^{-\i\vec{p}\cdot\vec{x} }  \Projector_{\overline{\gamma}\gamma} \langle N_\gamma(x) \overline{N}_{\overline{\gamma}}(0) \rangle \nonumber\\ 
&=&V_3 \volfac \sum_{\vec{x} \in V_3} \e^{-\i\vec{p}\cdot\vec{x} } \B(U(x,0),U(x,0),D(x,0)) \,,
\end{eqnarray}
where $\Projector$ is a matrix in Dirac space, $N_\gamma$ is an interpolating operator with the quantum numbers of a nucleon, and $U$ and $D$ are propagators corresponding to the $u$ and $d$ quarks respectively. We follow the conventions of \cite{Leinweber1991} and define
\begin{eqnarray}
\B(U_1,U_2,D)&=& \epsilon^{abc}\epsilon^{a'b'c'} \Tr \bigl \{ \Projector U_1^{aa'} \Tr_D [ (C\gamma_5)^T D^{cc' T} (C\gamma_5)  U_2^{bb'} ]   \nonumber\\
&+ & \Projector U_1^{aa'}  (C\gamma_5)^T D^{cc' T} (C\gamma_5)  U_2^{bb'}  \bigr \}\,, \label{eqn::contractions::2pt}
\end{eqnarray}
where the indices $a,b,c \in {1,2,3}$ refer to color.
A three-point function has a current insertion at one of the propagators:
\begin{align}
C^{3pt}_{\overline{N}N\Projector} (\tsink,\tins;\vec{p}_f, \vec{q},\Gamma, f) = V_3 \volfac\sum_{\vec{x} \in V_3} \e^{-\i\vec{p}_f\cdot\vec{x} } \Projector_{\overline{\gamma}\gamma} \langle N_\gamma(x) \mathcal{J}_f(\tins,\vec{q}, \Gamma ) \overline{N}_{\overline{\gamma}}(0) \rangle \,.
\end{align}
In this work we consider currents of the form
\begin{equation}
\mathcal{J}_f(\tins=y_4,\vec{q}, \Gamma )= \volfac\sum_{\vec{y} \in V_3} \e^{\i\qdoty} \Psibar_{f} \,^a_\alpha(y) \Gamma_{\alpha \beta} \Psi_{f} \,^a_\beta(y)\,,
\end{equation}
where $\Gamma$ is a matrix in Dirac space and $\Psi_{f}$ is a quark field of flavor $f$. We only consider connected contributions in this work.
The current insertion does not alter the Dirac and color structure of a propagator and the contractions for the three-point function  with a current of flavor $d$ yield

\begin{align}
C^{3pt}_{\overline{N}N\Projector} (\tsink,\tins;\vec{p}_f, \vec{q}, \Gamma, d)= V_3 \volfac \sum_{\vec{x} \in V_3} \e^{-\i\vec{p}_f\cdot\vec{x} } \B\left(U(x,0),U(x,0),\volfac\sum_{\vec{y} \in V_3} \e^{\i\qdoty} D(x,y)\Gamma D(y,0)\right). \label{eqn::contractions::3pt}
\end{align}
It is therefore possible to factorize the computation of a three-point function into a weight factor $W$ and a modified two-point function $\tilde{C}_{2pt}$ with the use of a timeslice-to-all propagator.

We can recover the three-point function from
\begin{align}
C^{3pt}_{\overline{N}N\Projector} (\tsink,\tins;\vec{p}_f, \vec{q}, \Gamma, d) =\frac{1}{N}\sum_{n=1}^N  \sum_{a,\alpha} \tilde{C}_{2pt}^{(n), d}(\tsink=x_4, \vec{p}_f)^a_{\alpha}W^{(n),a}_{\alpha}(d,d, \Gamma, \vec{q},\tsink= x_4,\tins= y_4 ) \label{eqn::stoch_three-pt_d}\,.
\end{align}

The timeslice-to-all propagator is computed by seeding the sink time slice with $(\mathbb{Z}_2 \times \i \mathbb{Z}_2)/\sqrt{2} $ source vectors $\eta^{(n)}_{\alpha,a} (x)$ \cite{Dong1994}. 

Sink vectors $\chi_f$, corresponding to the flavor $f$, are obtained by solving 
$\chi^{(n)}_{f,\beta,b}(y)=\mathcal{D}^{-1}_{f}\phantom{.}^{ba}_{\beta\alpha} (y,x) \eta^{(n)}_{\alpha,a} (x)$.
We can reconstruct the propagator from
\begin{equation}
\frac{1}{N}\sum_{n=1}^N \chi^{(n)}_{f,\beta,b}(y)\eta^{(n)\dagger}_{\alpha,a}(x)=G_{f, \beta \alpha}^{ba}(y,x)  \left( \Eins +\mathcal{O}\left(\frac{1}{\sqrt{N}}\right)\right) \label{eqn::AllToAllProp_definition} \,. 
\end{equation}
Using $\gamma_5$-hermiticity to revert the propagation direction, the weight factor is 
\begin{align}
W^{(n),a}_{\alpha}(f_1,f_2, \Gamma, \vec{q},\tsink= x_4,\tins= y_4 ) = \volfac \sum_{\vec{y} \in V_3} \e^{\i \qdoty } G_{f_1\alpha\beta'}^{aa'}(0,y) \Gamma_{\beta'\alpha'}(y) \chi^{(n)\dagger}_{f_2,\alpha',a'}(y)\gamma_5\,.
\end{align}
Note that the $\tsink$-dependence is implicit through the computation of $\chi$.
We define a pseudo propagator for the  $\tilde{C}_{2pt}$ as 
\begin{align}
\tilde{G}_{\tilde{\alpha}\beta}^{(n)\tilde{a}b}( x,0, a, \alpha) = \gamma_5\eta^{(n)}_{\beta,b}(x) \delta_{\tilde{a}a} \delta_{\tilde{\alpha}\alpha} \,.
\end{align}
 In the case of the $d$-quark current insertion in the proton the pseudo two-point function is
\begin{align}
\tilde{C}_{2pt}^{(n), d}(\tsink=x_4, \vec{p}_f)^a_{\alpha} =  V_3 \volfac \sum_{\vec{x} \in V_3} \e^{-\i\vec{p}_f\cdot\vec{x} } \B\left(U(x,0),U(x,0),\tilde{G}^{(n)}(x,0, a, \alpha)\right)\,.
\end{align}
The computation for the $u$-quark insertion is done in full analogy, the weight term is the same as the $u$ and $d$ quarks are taken to be degenerate in mass.
With this trick we have split up the double sum (insertion, sink) into a product of two sums which speeds up the calculation greatly as the current insertion is truly independent of the baryon sink (see \fig{pic:Comic}).

Matrix elements are computed by taking appropriate ratios of two- and three-point functions. Then the form factors in the Lorentz structure decompositions of the matrix elements are extracted by solving an overdetermined system of equations separately for each virtuality.
\newpage
\section{T-symmetry partner averaging}
\begin{wrapfigure}{r}{0.5\textwidth}
 \includegraphics[width=0.5\textwidth]{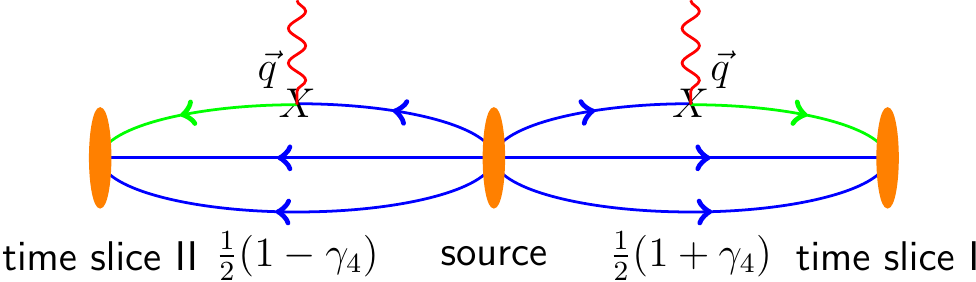}
 \caption{Pictorial representation of the double seeded stochastic three-point function.}
 \label{fig:Computation_fwd_backward} 
\end{wrapfigure}
Under time-reversal $t/a \rightarrow ( N_t - t/a )$ mod $N_t \equiv -t/a$, $\gamma_4 \rightarrow -\gamma_4$ the physics should not change. With a given point-to-all propagator we can compute the two-point functions of the nucleon using a positive parity projector $\Projector_+$ and a negative parity projector $\Projector_-$ separately at little additional cost. 
\begin{wrapfigure}{r}{0.5\textwidth}
 \includegraphics[width=0.5\textwidth]{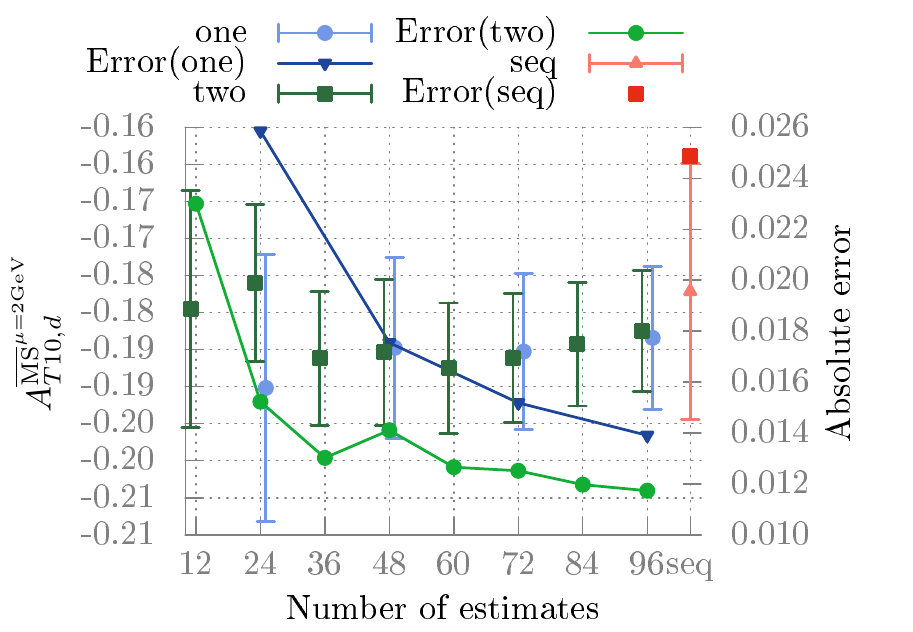}
  \caption{Comparison of single seeded to double seeded stochastic estimation method at $\beta=5.29$, $m_\pi\approx917$ GeV, $16^3\times 32$, $N_f=2$, Wilson Clover. The values for $A_{T10}$ were slightly shifted to ease the comparison. The lines indicate the absolute errors of both methods. For reference the results of the sequential source method are given.}
 \label{fig:GTb_plot}
\end{wrapfigure}
After time reversal of the negative parity projected two-point function the results can be averaged to increase the signal to noise ratio.

For the three-point function a similar procedure can be used. Since we then need a three-point function at $\tsink$ and another one at $-\tsink$ this would require additional inversions using the sequential source method. 
With the stochastic estimation we can seed the timeslices $\tsink$ and $-\tsink$ simultaneously. 
Then one can use the stochastic propagator to compute both three-point functions, see \fig{fig:Computation_fwd_backward}.
This increases the stochastic noise of each individual three-point function, but the total signal to noise ratio is improved, compared to having only one measurement per configuration.

We compute the ratios $C_{3pt}/C_{2pt}$ of the forward and backward propagating nucleon separately and then average as this keeps more of the correlations between two- and three-point function intact, compared to averaging first and then computing the ratio.
\section{Lattice parameters}
We have used  $N_f=2+1$, $32^3\times64$ SLiNC \cite{Horsley2008,Horsley2008a}  configurations from QCDSF at the symmetric point ($m_{u/d}=m_s$).
For the quark smearing we have used 400 steps of Wuppertal smearing \cite{Gusken1989} on APE-smeared \cite{Falcioni:1984ei}, gauge links. 
We have performed a study on a small sample (103 configurations) comparing the non-relativistic \cite{GoeckelerNucl.Phys.Proc.Suppl.42:337-3451995} sequential source method and the fully relativistic stochastic method. 
Omitting the anti-particle contributions of the di-quark contained in the sequential source reduces the computational effort by a factor of two 
and seems to be reliable in terms of groundstate overlap at this large pion mass.

The number of stochastic estimators was put to 60, which is a sweet spot in terms of signal/noise at fixed cost. 
This can be seen in \fig{fig:GTb_plot} where we plot the results for $A_{T10,d}$ \cite{Hagler2009} versus the number of estimates we have used. Employing a two timeslice source is compared to using only a one timeslice source and averaging the independently computed forward and backward going three-point function. 
For the stochastic method we average all three polarization directions, whereas for the sequential method we use only one. 
To extract all the form factors one needs 4  sequential sources (two flavors, polarized and unpolarized).
Neglecting other costs the non-relativistic sequential source method is comparable to 24 stochastic estimators (48 for the fully-relativistic sequential source).
Therefore, break-even is achieved if our errors are reduced by factors of 1.6 or 1.1, respectively.
\section{Results}
We have restricted the results to $Q^2 < 1$ GeV$^2$ as we do not claim to have reliable data above that threshold (the signal to noise on the individual ratios  deteriorates at too high momentum transfers). The stochastic data was furthermore restricted to use only $\vec{p}^{\,2}_f,\vec{p}^{\,2}_i  < 2 \frac{(2\pi)^2}{L^2}$ to keep the plots lucid.
To guide the eye we have also included the results for the stochastic method on 827 configurations. 

\begin{figure}[h]
\centering
 \includegraphics{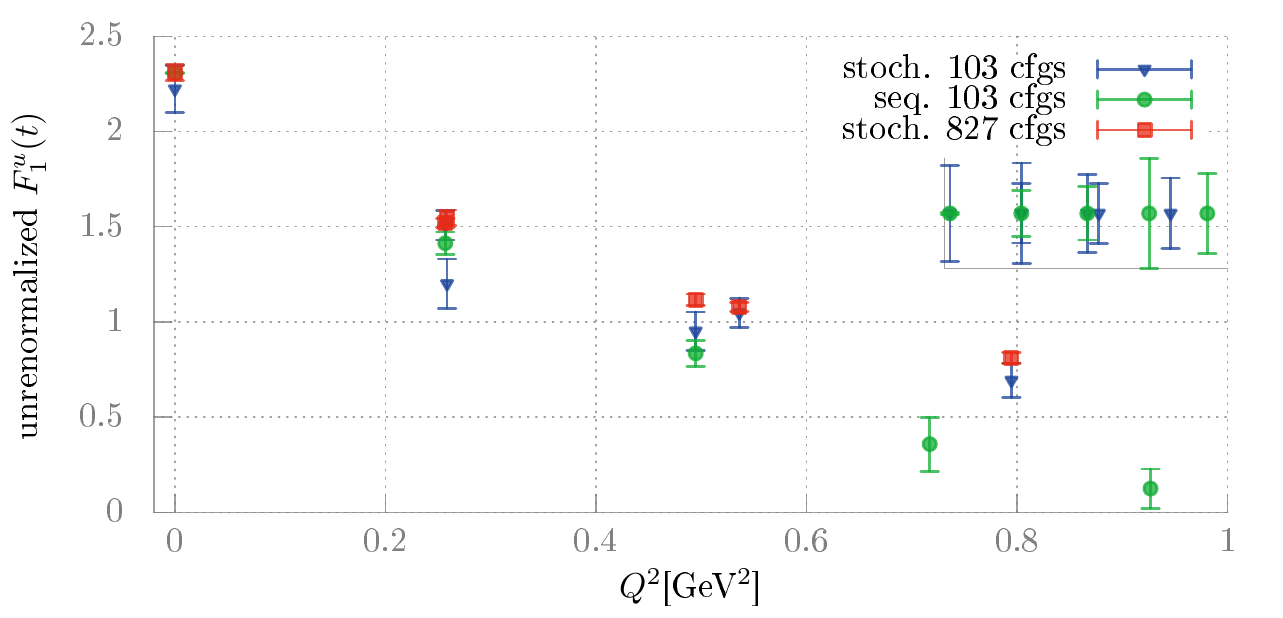}
\caption{Stochastic estimation vs. sequential source method for the electro-magnetic form factor $F_1$. In the inlay the results of the sequential and stochastic method for the small sample are shown with their values shifted to a constant. }
\label{fig::F1_stoch_vs_seq_erros}
\end{figure}

\begin{figure}[h]
\centering
 \includegraphics{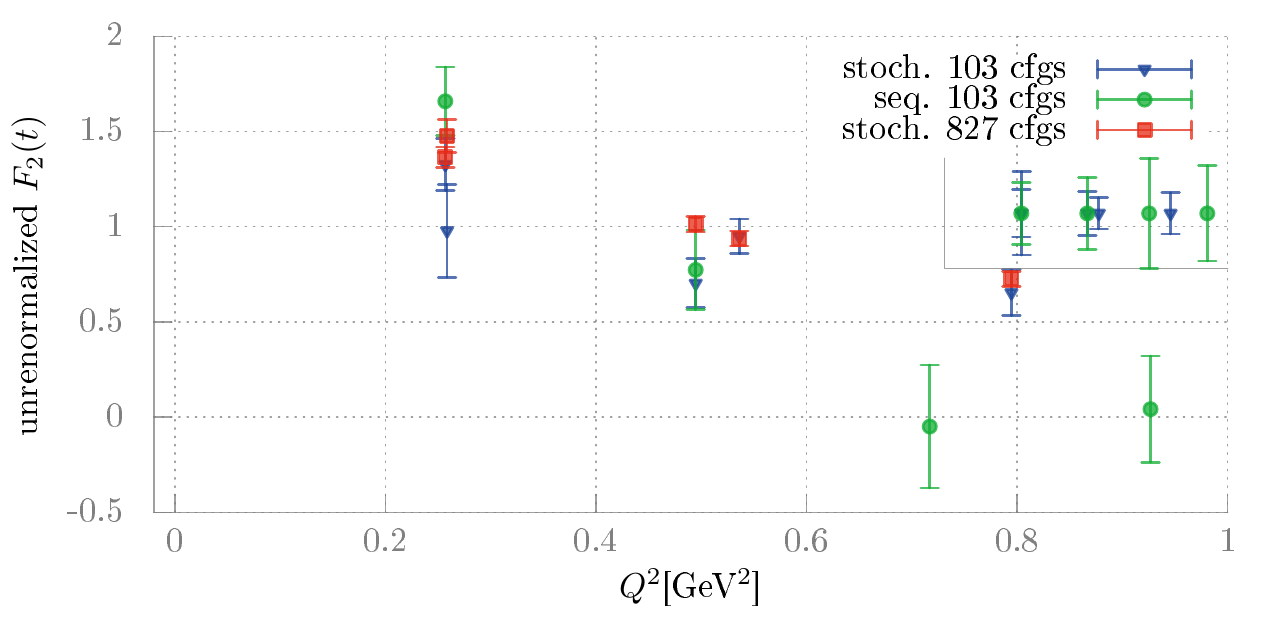}
\caption{Same as \protect\fig{fig::F1_stoch_vs_seq_erros} for the Pauli form factor $F_2$}
\label{fig::F2_stoch_vs_seq_erros}
\end{figure}

\begin{figure}[h]
\centering
\includegraphics{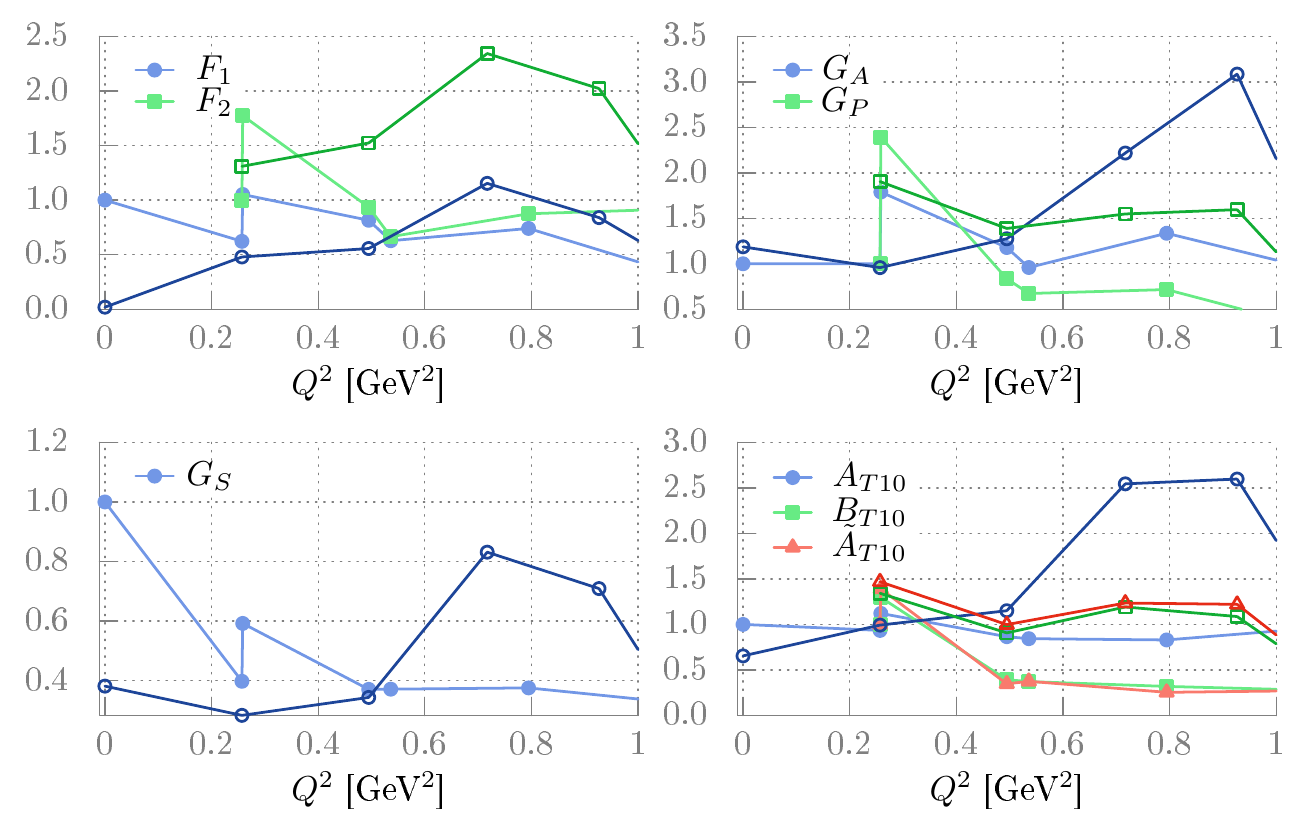}
\caption{Comparison of the errors of stochastic estimation (full symbols) vs. sequential source (open symbols) for the unrenormalized charge form factors. The errors are normalized with respect to the first defined value of the stochastic error in the channel.  }
\label{fig::Overview_stoch_vs_seq_errors_FF}
\end{figure}

For the electro-magnetic form factor $F_1$ (Dirac form factor), \fig{fig::F1_stoch_vs_seq_erros}, we see that the sequential method outperforms the stochastic method at low $Q^2$, but the higher number of ratios that can be evaluated with the latter method allows us to determine $F_2$ (the Pauli form factor) much better already at medium $Q^2$, see \fig{fig::F2_stoch_vs_seq_erros}. 
For the other form factors,  especially those with many contributions from polarized nucleons, the performance of the stochastic method improves even more significantly towards high $Q^2$. 
This is illustrated in \fig{fig::Overview_stoch_vs_seq_errors_FF}. 
Note that the stochastic method provides two measurements at two very similar virtualities (e.g., $\vec{p}_i=(1,0,0)\rightarrow p_f=(0,0,0)$ and $\vec{p}_i=(1,1,0)\rightarrow p_f=(1,0,0)$) due to the sink momentum not being restricted to zero.

\section{Conclusions and outlook}
The stochastic method to compute baryon three-point functions is competitive to the sequential source method and is more flexible enabling the computation of multiple $\tsink$, sink smearings, or even a realizing different baryons using the same stochastic propagator. This method is therefore particularly suitable for computing the whole set of three-point functions of the baryon octet/decuplet with transitions.

\section{Acknowledgements}
We use the Chroma software suite \cite{EdwardsNucl.Phys.Proc.Suppl.140:8322005} and part of the calculation was performed on JuRoPA at the J\"ulich Supercomputer Center, and iDataCool in Regensburg. This work was supported by SFB-TR55(Hadron Physics from Lattice QCD) and the EU ITN STRONGnet (238353).

\providecommand{\href}[2]{#2}\begingroup\raggedright\endgroup

\end{document}